\documentclass[prb,aps,showpacs,prb,amsmath,floatfix,showkeys]{revtex4}
\usepackage[]{graphicx}
\usepackage{graphicx}% Include figure files
\usepackage{dcolumn}% Align table columns on decimal point
\usepackage{bm}% bold math

\def \bea { \begin{eqnarray} }
\def \be { \begin{equation} }
\def \ee { \end{equation} }
\def \eea { \end{eqnarray} }

\begin{document}

\title{Tunable Fano resonance in a parallelly coupled diatomic 
molecular transistor}

\author{A. Goker}

%\affiliation{$^2$
%Department de Chimie, Universite de Montreal, \\
%C.P.6128 Succursale A, Montreal, Quebec H3C 3J7, Canada
%}

\affiliation{
Department of Physics, Fatih University, \\
Buyukcekmece, Istanbul 34500, Turkey
}

\date{\today}

\begin{abstract}
We investigate electron transport through a diatomic molecule parallelly
coupled to infinite source and drain contacts. We utilize a model Hamiltonian involving
a Hubbard term in which the contacts are modeled using recently developed complex
source and sink potentials. The zero bias transmission spectrum for a symmetrically 
coupled system as a function of the Fermi energy acquires a Fano lineshape as the Hubbard 
interaction is turned on. For large values of $U$, the Fano lineshape broadens
and shifts to higher energy values disappearing eventually. Meanwhile, the Breit-Wigner
resonance located at the bonding resonance in the noninteracting limit survives but its 
position is shifted twice the coupling between the atoms in the molecule in the infinite $U$ 
limit and its linewidth is reduced to half. We attribute this behaviour to the unavailability of 
one of the transmission channels due to Coulomb blockade.     
\end{abstract}
\maketitle

Molecular electronics \cite{JGA00,N01,NR03,HR03,CHH07} was
put forward as a novel paradigm for the construction of future electronic
devices almost three decades ago. \cite{AvirametalCPL74} However, its feasibility has been 
much debated since it seemed to offer no clear alternative to 
the conventional bits for the transmission of information. The advent of spintronics 
\cite{Zuticetal04RMP,Bratkovsky08RPP}, where spin and charge transport are 
investigated in mesoscopic systems, provided a satisfying resolution to this controversy 
by proposing to utilize the spin degrees of freedom of the electron as bits.  
These developments culminated in the emergence of the field of molecular spintronics 
which aims to incorporate spin dependent transport into molecular electronics.
\cite{Rochaetal05,Seneoretal07JPCM}. 

Effective one-electron theories based on a combination of 
non-equilibrium Green's function technique and density functional theory 
have dominated the theoretical analysis of molecular
electronic devices (MED's) \cite{MKR94,D95,NDL95,XDR01,TGW01a,BMO02,EZ05}. 
However, these models fail to account for electron correlation effects 
\cite{EZ05,kbe06} since the choice of an exchange-correlation functional which
can capture the self-energy of the electron adequately is unknown. 
This is reflected in large discrepancies between theoretical predictions and experimental
results. Therefore, it is of crucial interest to be able to develop 
theoretical methods that can tackle this problem.

Hedin's celebrated GW approximation \cite{HedinPR65,HedinJPCM99} has been invoked recently 
in conjunction with a model Hamiltonian to overcome the self-energy deficiencies of 
one-electron theories. \cite{ThygesenPRL08,ThygesenetalPRB08,ThygesenetalJCP07} 
For the same purpose, we implemented the source-sink potential method (SSP)
\cite{GEZ07,Ernzerhof07JCP} in a model Hamiltonian involving a Hubbard term to probe 
the strongly correlated regime in molecular transistors systematically. 
In the case of a diatomic molecule serially coupled to ferromagnetic contacts, 
we found that the electron transport is suppressed in the $U \rightarrow \infty$ limit 
while the energy gap between the bonding and antibonding resonances is reduced 
due to a coupling between the molecular singlet and triplet states for small $U$. 
\cite{Gokeretal08JCP}
 
In this paper, we extend our analysis and analyze a diatomic
molecular transistor parallelly coupled to infinite ferromagnetic contacts. Starting
from the known results in noninteracting limits, we explore the transmission of the
system in infinitesimal bias as a function of the Fermi energy of the contacts for
various values of the intrasite interaction strength $U$. Our results indicate that turning 
on the interaction results in appearance of a Fano resonance  when the molecule is symmetrically 
coupled to contacts. Increasing the interaction strength produces several effects.
The Breit-Wigner resonance located at the bonding level in the noninteracting limit
starts moving towards the antibonding level. During this shift, it transfers some of its
linewidth to the Fano resonance which also starts moving towards higher energies.
Hence, sweeping the interaction strength provides tunability for the Fano resonance.
Even though tunable Fano resonances have been reported for other systems ranging 
from metallic nanostructures consisting of a disk inside a thin ring \cite{HaoetalNL08}
to localized Bose-Einstein condensates in one-dimensional optical lattices \cite{VicencioetalPRL07}, 
this paper constitutes the first prediction of a tunable Fano resonance in a strongly 
correlated molecular transistor. 
In the $U \rightarrow \infty$ limit, the Fano resonance
disappears to high energies whereas the Breit-Wigner resonance survives by settling into the 
antibonding position of the noninteracting system with its linewidth reduced to half. 

The source-sink potential method relies on a Bloch wave impinging on the contact molecule
interface from the left contact. It is partially reflected and partially transmitted 
at the junction. The reflection coefficient is found by solving the Schr$\ddot{o}$dinger
equation for the relevant Hamiltonian and the zero bias transmission is obtained from
the Landauer-B$\ddot{u}$ttiker formula. In this approach, the semi-infinite contacts are described 
with a H$\ddot{u}$ckel model and  we assume that they are ferromagnetic with parallel configurations, 
i.e. they only contain spin-up electrons. This requirement can be realized experimentally.
\cite{Seneoretal07JPCM,Rochaetal05}

The atoms in the semi-infinite contacts are coupled to their nearest neighbour with 
strength $\beta_C$. The first and last atoms in the right and left contacts are each coupled 
to the molecular atoms through $\beta_{CM}$. The Hamiltonian is given by
\begin{equation}
H=h_{\uparrow}+h_{\downarrow}+V_{\uparrow\downarrow}
\label{medham}
\end{equation}
where
\begin{eqnarray}
h_{\uparrow}&=&\Sigma_L n_{L\uparrow}+\Sigma_R n_{R\uparrow}+
\beta_{CM}c^{\dagger}_{L\uparrow}c_{a\uparrow}+H.c.+ \nonumber \\
& & \beta_{CM}c^{\dagger}_{L\uparrow}c_{b\uparrow}+H.c.+\beta_{CM}c^{\dagger}_{R\uparrow}c_{a\uparrow}+H.c.+ \nonumber \\
& & \beta_{CM}c^{\dagger}_{R\uparrow}c_{b\uparrow}+H.c.+\beta_M c^{\dagger}_{a \uparrow}c_{b \uparrow}+H.c.
\end{eqnarray}
\begin{eqnarray}
h_{\downarrow}&=&\beta_{CM}c^{\dagger}_{L\downarrow}c_{a\downarrow}+H.c.+\beta_{CM}c^{\dagger}_{L\downarrow}c_{b\downarrow}
+H.c.+ \nonumber \\
& & \beta_{CM}c^{\dagger}_{R\downarrow}c_{a\downarrow}+H.c.+\beta_{CM}c^{\dagger}_{R\downarrow}c_{b\downarrow}+H.c.+\nonumber \\
& & \beta_M c^{\dagger}_{a \downarrow}c_{b \downarrow}+H.c.
\end{eqnarray}
and
\begin{equation}
V_{\uparrow\downarrow}=\sum_{\alpha} U n_{\alpha\uparrow}n_{\alpha\downarrow}.
\end{equation}
The operators $c^{\dagger}_{\alpha\sigma}$ ($c_{\alpha\sigma}$) with $\alpha$=$a$, $b$ and
$c^{\dagger}_{\beta\sigma}$ ($c_{\beta\sigma}$) with $\beta$=L,R create(destroy)
an electron in the molecular sites a and b and contacts L and R respectively.
The source $ \Sigma_L =\beta_C \frac{e^{-iq}+re^{iq}}{1+r} $ and sink $ \Sigma_R =\beta_C e^{iq} $
potentials \cite{GEZ07} appearing in the Hamiltonian, where $q=arccos\left(\frac{\epsilon}{2\beta_C}\right)$,
describe the infinite parts of single-electron contacts rigorously. \cite{GEZ07}
 
Our finite dimensional Hilbert space contains the isolated molecule configurations where
the spin up and down electrons are confined within the atoms of the molecule. This results 
in four states. Additionally, we include the configurations in which the up-spin electron 
from atom $ a $ or atom $ b $ is excited into the contacts. This gives us 
another four states. A detailed description of how we obtain these states
has been discussed elsewhere. \cite{Gokeretal08JCP} This model is presumably the 
simplest realization of a molecule interacting with ferromagnetic contacts.

The Hamiltonian in Eq.~\ref{medham} can then be represented in the space of the configurations as
\[  \left( \begin{array}{llllllll}
\Sigma_L & \beta_M  &  \frac{\beta_{CM}}{\sqrt{2}} &  \frac{-\beta_{CM}}{\sqrt{2}} & \beta_{CM} & 0 & 0 & 0 \\
\beta_M  & \Sigma_L &  \frac{\beta_{CM}}{\sqrt{2}} & \frac{\beta_{CM}}{\sqrt{2}} & 0 & \beta_{CM} & 0 & 0 \\
\frac{\beta_{CM}}{\sqrt{2}} & \frac{\beta_{CM}}{\sqrt{2}} & 0 & 0 & \sqrt{2}\beta_{M} & \sqrt{2}\beta_{M} & \frac{\beta_{CM}}{\sqrt{2}} 
& \frac{\beta_{CM}}{\sqrt{2}} \\
\frac{-\beta_{CM}}{\sqrt{2}} & \frac{\beta_{CM}}{\sqrt{2}} & 0 & 0 & 0 & 0 & \frac{-\beta_{CM}}{\sqrt{2}} & \frac{\beta_{CM}}{\sqrt{2}} 
\\
\beta_{CM} & 0 & \sqrt{2}\beta_{M} & 0 & U & 0 & \beta_{CM} & 0 \\
0 & \beta_{CM} & \sqrt{2}\beta_{M} & 0 & 0 & U & 0 & \beta_{CM} \\
0 & 0 & \frac{\beta_{CM}}{\sqrt{2}} & \frac{-\beta_{CM}}{\sqrt{2}} & \beta_{CM} & 0 & \Sigma_R & \beta_M \\
0 & 0 & \frac{\beta_{CM}}{\sqrt{2}} & \frac{\beta_{CM}}{\sqrt{2}} & 0 & \beta_{CM} & \beta_M & \Sigma_R \\
\end{array} \right),\]

The procedure to obtain the transmission probability is straightforward once we are equipped with 
the appropriate Hamiltonian. It has been described in detail previously. Here we just summarize the 
method briefly. The transmission probability $T(\epsilon)$ describes the probability of an incoming 
electron with Fermi energy $ \epsilon $ to penetrate the molecule. It is related to the reflection 
coefficient $r(\epsilon)$, which is buried within the 
potential $\Sigma_L$ inside the Hamiltonian, by $T(\epsilon)=1 -|r(\epsilon) |^2$.

The Schr$\ddot{o}$dinger equation involving the Hamiltonian given above becomes
\be
H(\epsilon,r) \Psi = ( E^{N-1} + \epsilon ) \Psi,
\label{eigen2}
\ee
where $ E^{N-1} $ is the energy of the  $N-1$-electron system resulting from removing the up-spin electron.
\cite{KD80,EBP96} In this paper, we are concerned about the evolution of the molecular ground state, thus we set
$ E^{N-1}=\beta_M $ where $ \beta_M $ is the energy of the bonding orbital in the isolated molecule.

The corresponding eigenvalue equation turns out to be
\be
Det (H(\epsilon,r) - (\beta_M + \epsilon )I)=0.
\label{eigenvalue}
\ee
Eq.~\ref{eigenvalue} determines the reflection coefficient $ r $ in terms
of the variable $ \epsilon $ for a given $ \beta_M $. We solve this
secular equation for successive values of the Fermi energy $ \epsilon $
to obtain the entire transmission spectrum. 

\begin{figure}[h]
\includegraphics[width=6.5cm]{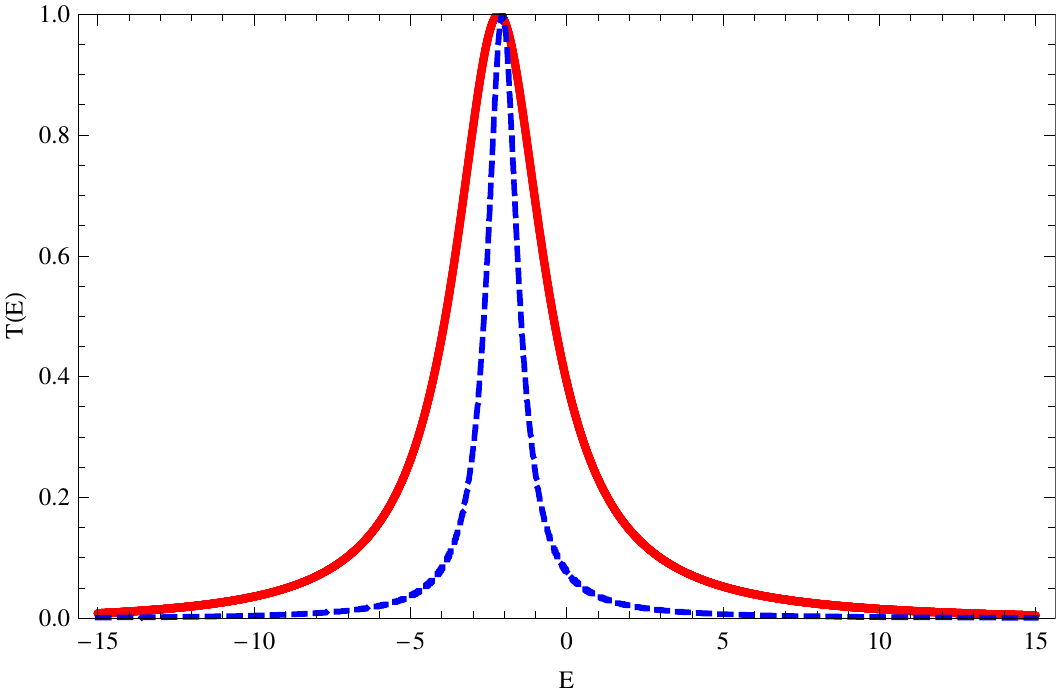}
\caption{The transmission probability $T(\epsilon)$ for a non-interacting
system as a function of Fermi energy $\epsilon$. Red(solid) curve
corresponds to $ \beta_C=-10 $, $ \beta_M=\beta_{CM}=-2 $ whereas
blue(dashed) curve is obtained for  $ \beta_C=-10 $, $ \beta_M=-2 $ and $ \beta_{CM}=-1 $.}
\label{noninteracting}
\end{figure}

We can now embark on our analysis of the transmission spectrum. We start with 
the noninteracting limit. 
This will help us to verify the validity of our approach and elucidate the
implications of the choice of the parameters since the transmission
results in this limit have already been reported.\cite{Guevaraetal03PRB}
Fig.~\ref{noninteracting} depicts the transmission spectrum for 
a noninteracting system, i.e. U=0, for two different values of the coupling to
the contacts $ \beta_{CM} $. We consider a system in which the atoms in the
contacts are tightly coupled compared to the intermolecular and molecule-atom
coupling therefore we pick $ \beta_C $ much larger than the other two couplings.
This choice has been motivated by the fact that it enables us to analyze a wider
transmission spectrum.

\begin{figure}[h]
\includegraphics[width=6.4cm]{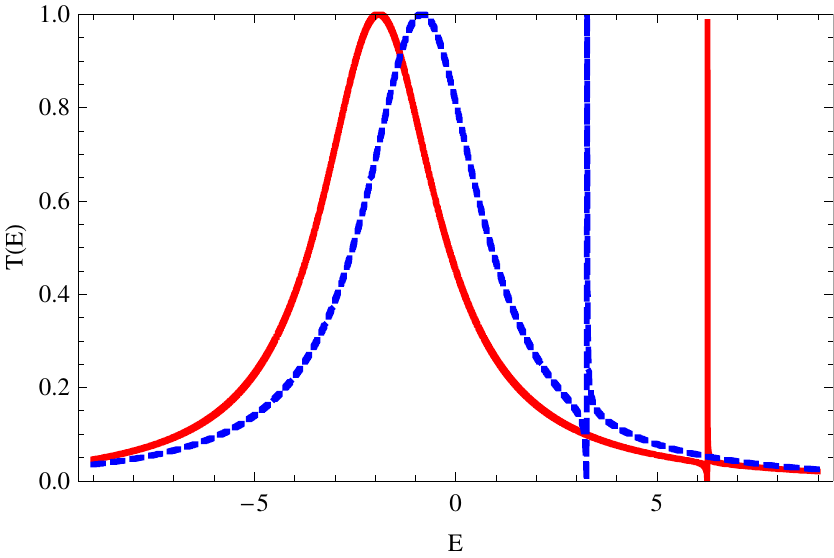}
\caption{The transmission probability $T(\epsilon)$ for an interacting  
system with $U$=0.5 as a function of Fermi energy $\epsilon$. Red(solid) curve
corresponds to $ \beta_C=-10 $, $ \beta_M=\beta_{CM}=-2 $ whereas
blue(dashed) curve is obtained for  $ \beta_C=-10 $, $ \beta_M=-1 $ and $ \beta_{CM}=-2 $.}
\label{weak}
\end{figure}

A glance at this figure reveals that the spectrum consists of nothing but 
a Breit-Wigner resonance located at the bonding energy of the isolated molecule with 
linewidth equal to 2 $ \mid \beta_{CM} \mid $. Therefore, our results are in full agreement
with the previously reported data. \cite{Guevaraetal03PRB}
The disappearance of the Fano resonance at the antibonding energy is a result of the
emergence of slow transitions between the antibonding state and the contacts. 
Moreover, the investigation of the local density of states in each atom
within the molecule shed light on this peculiarity by showing that 
the lorentzian Breit-Wigner resonance located at the antibonding resonance 
becomes a dirac delta function at symmetric coupling. Consequently, the antibonding 
state gets completely localized inside the molecule. \cite{Guevaraetal03PRB}
 
After confirming the earlier results in the noninteracting limit with our model,
we are now ready to start exploring the strongly correlated regime. For this purpose,
we gradually turn on the intraatomic Coulomb repulsion within the molecule.
Fig.~\ref{weak} shows the transmission spectrum for two configurations
with different interatomic couplings in the weakly correlated regime. The noteworthy
feature here is the appearance of a sharp Fano resonance alongside with the Breit-Wigner
resonance. Fano resonance is an exquisite interference effect arising from the existence
of two transport pathways, a resonant one and a nonresonant one.\cite{FanoPR61} As a result, it
is quite sensitive to the phase coherence of the system. The Fano line shape is 
given by
\be
G(\epsilon)=G_d \frac{\mid 2(\epsilon-\epsilon_0)+q\Gamma \mid^2}{4(\epsilon-\epsilon_0)^2+\Gamma^2}
\ee
where $\Gamma$ is the linewidth of the resonance, $G_d$ is the nonresonant conductance and
$q$ is the Fano parameter. It is defined as the ratio between the resonant tunneling
probability and the direct nonresonant tunneling probability. The $ q \rightarrow \infty $ limit 
yields the Breit-Wigner resonance.  

Our transmission results bear close resemblance to the transport spectroscopy of double quantum dot
Aharonov-Bohm (AB) interferometers. An Aharonov-Bohm interferometer
is in essence a double-slit experiment in which an electron tunneling from the 
left contact to the right contact is split into two waves.\cite{KonigetalPRB02} A magnetic flux $ \Phi $ 
piercing the area enclosed by the two paths induces a change in the relative phase of
the amplitudes of the two waves. The resulting transmission spectrum has been shown to exhibit 
a Breit-Wigner and a Fano resonance in the noninteracting limit.\cite{KangetalJPCM04} The most 
striking part of our result is that we can achieve the same transmission spectrum in symmetric 
coupling even in zero magnetic field. The Hubbard interaction inside the molecule is responsible 
for the decoherence in our case instead of the applied external magnetic field in AB interferometers. 
More recently, taking into account the spin-orbit coupling in a noninteracting Rashba quantum dot 
coupled to ferromagnetic contacts has also been shown to result in a transmission spectrum
composed of a convolution of a Fano lineshape and A Breit-Wigner lineshape.\cite{OrellanaetalNano08}

One particular intricacy that deserves further attention in Fig.~\ref{weak} is the separation
between the Fano and Breit-Wigner resonances. It is clear from this figure that the location
of the Fano resonance is pushed to higher energies as the coupling between the atoms
inside the molecule increases. In fact, when $\mid \beta_M \mid$ exceeds twice $\mid \beta_{CM} \mid$, the
Fano resonance is no longer visible in the transmission spectrum even for very small $U$ values.
This makes perfect sense intuitively because these conditions mean that the 
tunneling rate of the down-spin electron between the molecular atoms is much higher
than the tunneling rate of the up-spin electrons from the contacts to the molecule or
vice versa. Therefore, the existence of two distinct pathways for the up-spin
electron gradually disappears leading to the elimination of the Fano resonance from
the transmission spectrum. 

\begin{figure}[h]                                                                      
%\centerline{\includegraphics[width=4.5cm]{Newfig3a.pdf}                                   
%\includegraphics[width=4.5cm]{Newfig3b.pdf}}
\includegraphics[width=6.3cm, height=4.3cm]{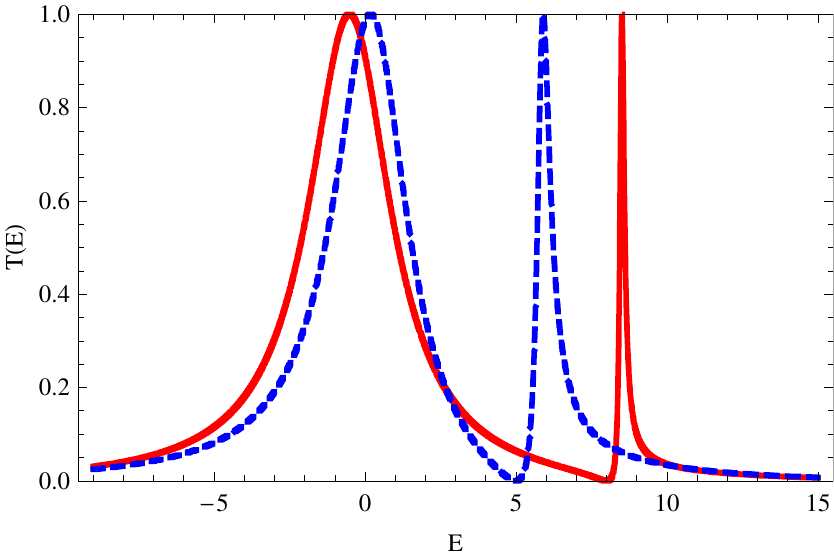}
\includegraphics[width=6.3cm, height=4.3cm]{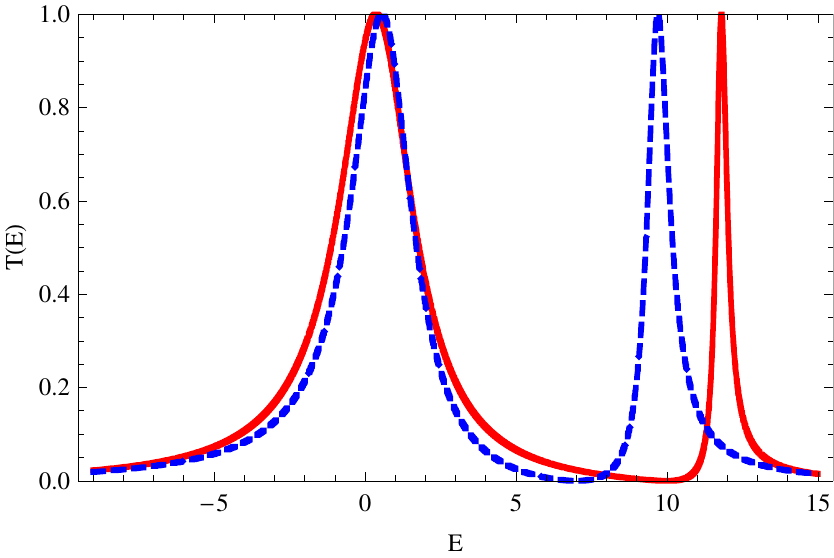}
\caption{Left panel shows the transmission probability  $T(\epsilon)$ 
for an interacting system with $U$=4 as a function of the
Fermi energy $\epsilon$. Red(solid) curve
corresponds to $ \beta_C=-10 $, $ \beta_M=\beta_{CM}=-2 $ whereas
blue(dashed) curve is obtained for  $ \beta_C=-10 $, $ \beta_M=-1 $ and $ \beta_{CM}=-2 $.
The transmission spectrum in the right panel is obtained 
by using the same couplings in left panel for $U$=8.
}
\label{medium}
\end{figure}

We keep increasing the strength of the electron-electron interaction inside
the molecule. The transmission results for $U$=4 and $U$=8 are depicted in the
left and right panels in Fig.~\ref{medium} respectively. The implications of
this are two fold. First, both the Breit-Wigner resonance and the Fano resonance
start shifting to higher energies. Meanwhile, the linewidth of the Breit-Wigner
resonance starts decreasing whereas the Fano resonance gets broadened. We checked
that the total linewidth stays constant. This suggests that the Breit-Wigner
resonance is transfering some of its linewidth to the Fano resonance as $U$ 
starts increasing. This is simply because the direct nonresonant electron
transport occurring through one of the channels where the down-spin electron is
located is becoming less likely as the Coulomb repulsion is pushing
the up-spin electron away from it. Consequently, the Fano resonance, which
is arising from the interference of the nonresonant and resonant channels,
is acquiring the lost linewidth in the Breit-Wigner resonance.

\begin{figure}[h]
\centerline{\includegraphics[width=6.4cm]{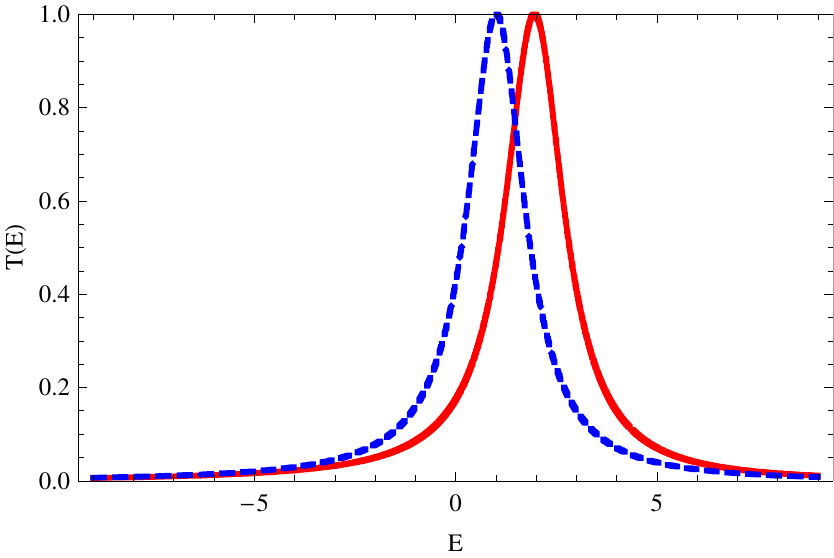}}
\caption{The transmission probability $T(\epsilon)$ for an interacting
system with $U$=128 as a function of Fermi energy $\epsilon$. Red(solid) curve
corresponds to $ \beta_C=-10 $, $ \beta_M=\beta_{CM}=-2 $ whereas
blue(dashed) curve is obtained for  $ \beta_C=-10 $, $ \beta_M=-1 $ and $ \beta_{CM}=-2 $.}
\label{strong}
\end{figure}

The transmission results in the strongly correlated limit are depicted in Fig.~\ref{strong}.
Increasing the value of $U$ any further does not alter the shape or the location of
the resonances therefore we are content to show the results for only $U$=128.
It is apparent from this figure that the Fano resonance has been pushed to higher energies
hence it is no longer visible. Moreover, the Breit-Wigner resonance in the noninteracting
limit has shifted 2$\mid \beta_M \mid$ to settle into the antibonding energy while its linewidth has been 
reduced to half. This behaviour is entirely different from the serially coupled case where the electron transport is
completely inhibited due to the Coulomb blockade of the only transport channel. \cite{Gokeretal08JCP}
In parallel coupling, there are two distinct transport channels and even though the
one containing the down spin electron is now unavailable due to Coulomb blockade, the
other one still permits hopping. We believe this is precisely why the linewidth of
the Breit-Wigner resonance has been halved compared to the noninteracting case. Moreover,
the fact that the electron transport has shifted to the antibonding energy at this limit
shouldn't come as a surprise. An up-spin electron hopping from left contact
to the right contact uses only the atom in which the down-spin electron is not located
without making a detour to the atom where the down-spin electron is residing.
Therefore, intramolecular hopping is largely avoided in this regime resulting in
non-overlapping wavefunctions for the atoms of the molecule. This is nothing but the
antibonding configuration.

In this paper, we investigated the electron transport through a diatomic molecular electronic 
device parallelly coupled to infinite ferromagnetic contacts at zero temperature. We modeled the system 
by using a model Hamiltonian involving Hubbard term and the contacts were described with the aid of 
recently  developed source and sink potentials. Our results showed that the transmission spectrum 
exhibits a Fano resonance in the interacting regime as well as the usual Breit-Wigner resonance. 
We explained this as a result of the decoherence introduced by the Coulomb interaction with 
the electron residing inside the molecule. As the electron-electron 
interaction strength inside the molecule has been increased, we witnessed the gradual shift of 
both resonances to higher energies while the Breit-Wigner resonance transfered some of its linewidth
to the Fano resonance. When we approached the $ U \rightarrow \infty $ limit, only the Breit-Wigner
resonance survived and got shifted by twice the coupling between the atoms
in the molecule. We attributed this to the unavailability of the one of the transport
channels due to Coulomb blockade while the other one remains open. This conclusion is supported
by the reduction of the linewidth of the Breit-Wigner resonance to half of its noninteracting
value. We would like to point out that the scenario that has been discussed in this paper
can be realized experimentally. The value of the Hubbard interaction strength relative
to the other couplings can be adjusted by stretching the junction. We expect that the
$ U \rightarrow \infty $ limit would be attained when the bond is about to dissociate.

A.G thanks Prof. Barry Friedman for a critical reading of the manuscript and
acknowledges fruitful discussions with Prof. Matthias Ernzerhof.
%\clearpage

\bibliographystyle{iopams}
\bibliography{gen,ref}
\end{document}